\documentclass[aps,prl,preprint,superscriptaddress]{revtex4-1}

\usepackage{lineno}

\usepackage{graphicx}
\usepackage{epstopdf}
\usepackage[utf8]{inputenc}
\usepackage{amsmath}
\usepackage[T1]{fontenc}
\usepackage{lmodern}
\begin{document}


\title{Dynamical Negative Differential Resistance in Antiferromagnetically Coupled Few-Atom Spin-Chains}



\author{Steffen Rolf-Pissarczyk}
\thanks{These authors contributed equally to this work}
\email[]{steffen.rolf-pissarczyk@mpsd.mpg.de}
\affiliation{Max Planck Institute for the Structure and Dynamics of Matter, Luruper Chaussee 149, 22761 Hamburg, Germany }
\affiliation{Max Planck Institute for Solid State Research, Heisenbergstraße 1, 70569 Stuttgart, Germany}

\author{Shichao Yan}
\thanks{These authors contributed equally to this work}
\email[]{yanshch@shanghaitech.edu.cn}
\affiliation{Max Planck Institute for the Structure and Dynamics of Matter, Luruper Chaussee 149, 22761 Hamburg, Germany }
\affiliation{Max Planck Institute for Solid State Research, Heisenbergstraße 1, 70569 Stuttgart, Germany}
\affiliation{School of Physical Science and Technology, ShanghaiTech University, Shanghai 201210, China}

\author{Luigi Malavolti}
\affiliation{Max Planck Institute for the Structure and Dynamics of Matter, Luruper Chaussee 149, 22761 Hamburg, Germany }
\affiliation{Max Planck Institute for Solid State Research, Heisenbergstraße 1, 70569 Stuttgart, Germany}

\author{Jacob A.J. Burgess}
\affiliation{Max Planck Institute for the Structure and Dynamics of Matter, Luruper Chaussee 149, 22761 Hamburg, Germany }
\affiliation{Max Planck Institute for Solid State Research, Heisenbergstraße 1, 70569 Stuttgart, Germany}
\affiliation{Department of Physics and Astronomy, University of Manitoba, Winnipeg, Manitoba R3T 2N2 Canada}

\author{Gregory McMurtrie}
\affiliation{Max Planck Institute for the Structure and Dynamics of Matter, Luruper Chaussee 149, 22761 Hamburg, Germany }
\affiliation{Max Planck Institute for Solid State Research, Heisenbergstraße 1, 70569 Stuttgart, Germany}

\author{Sebastian Loth}
\email[]{sebastian.loth@mpsd.mpg.de}
\affiliation{Max Planck Institute for the Structure and Dynamics of Matter, Luruper Chaussee 149, 22761 Hamburg, Germany }
\affiliation{Max Planck Institute for Solid State Research, Heisenbergstraße 1, 70569 Stuttgart, Germany}
\affiliation{University of Stuttgart - Insitute for Functional Matter and Quantum Technologies, Pfaffenwaldring 57, 70569 Stuttgart, Germany}



\date{\today}

\begin{abstract}
We present the appearance of negative differential resistance (NDR) in spin-dependent electron transport through a few-atom spin-chain. A chain of three antiferromagnetically coupled Fe atoms (Fe trimer) was positioned on a Cu$_2$N/Cu(100) surface and contacted with the spin-polarized tip of a scanning tunneling microscope, thus coupling the Fe trimer  to one non-magnetic and one magnetic lead. Pronounced NDR appears at the low bias of 7 mV where inelastic electron tunneling dynamically locks the atomic spin in a long-lived excited state. This causes a rapid increase of the magnetoresistance between spin-polarized tip and Fe trimer and quenches elastic tunneling. By varying the coupling strength between tip and Fe trimer we find that in this transport regime the dynamic locking of the Fe trimer competes with magnetic exchange interaction, which statically forces the Fe trimer into the high-magnetoresistance state and removes the NDR. 

\end{abstract}


\maketitle

Spintronics utilizes the electron spin as a further degree of freedom in solid-state devices. 
New spintronic applications in magnetic sensing, data storage and quantum information processing are enabled \cite{Wolf2001,Zutic2004,Bogani2008} because of the low power needed to manipulate spins compared to charge-based electronics. due to their nonvolatility and low power consumption compared to the charge-based electronics. In this respect, there is great interest in translating general functionalities of conventional charge-based electronics into a spin-based footing. 
\par
Negative differential resistance (NDR), referring to a decrease in current with increasing voltage, is an important effect that enables a two-terminal device to operate as an amplifier or oscillator. It is usually achieved by the shift or occupation of electronic states in semiconducting tunnel junctions \cite{Esaki1974, Grundmann2006}, transport through atoms, molecules and quantum dots \cite{Chen1999, Perrin2014,Xu2015, Gaudioso2000,Tao2006, Warner2015}, or recently by the charging dynamics of dopants \cite{Rashidi2016}.
\par
Spin-based NDR has been reported for electron transport through single or coupled quantum dot devices \cite{Weinmann1995,Ono2002,Pioro-Ladriere2003} or through single molecule magnets involving nondegenerate spin multiplets \cite{Heersche2006}.  It has been proposed that spin-based NDR, purely based on spin-spin interaction, will appear in inelastic electron transport through anisotropic magnetic systems coupled to one ferromagnetic lead and one nonmagnetic lead \cite{Elste2006} but this effect has hitherto not been observed experimentally.    
\par

Here we report the appearance of spin-based NDR in inelastic electron transport through an anisotropic antiferromagnetically coupled few-atom spin-chain in a spin-polarized scanning tunneling microscope (STM). In this configuration, the spin-polarized tip acts as the magnetic lead with a precisely adjustable position, and the sample substrate as the nonmagnetic lead which the spin-chain directly lies on. Inelastic scattering of electrons locks the  spin-chain into an excited spin state with large magnetoresistance. This suppresses elastic tunneling between ferromagnetic lead and  spin-chain causing NDR at very low voltage. By varying the STM tip-sample distance we adjust the coupling strength between spin-polarized tip and spin-chain, and measure variations in the NDR. We find that electron tunneling competes with magnetic exchange interaction between the spin-polarized tip and  spin-chain that occurs when bringing the  spin-chain into contact with the spin-polarized tip. For weak coupling, the dynamic locking of the  spin-chain produces prominent NDR, whereas for strong coupling, the magnetic exchange interaction removes the NDR by forcing the  spin-chain into a spin state configuration that does not permit it anymore.
\par

 \begin{figure}
 \includegraphics{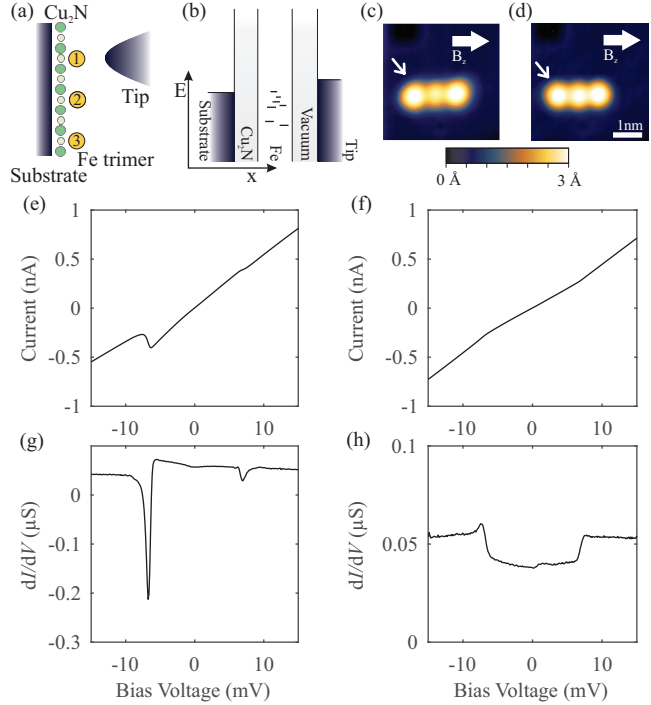}
 \caption{(a) Schematic of the experimental setup. (b) The Cu$_2$N layer and the vacuum form a two barrier structure around the Fe~trimer.  (c) and (d) STM constant current topographies of a few-atom  spin-chain measured with a spin-polarized tip (c) and  a non-spin-polarized tip (d) at a tunnel junction setpoint of 2 nS with 5 mV. (e) and (f) $I(V)$ curves recorded a spin-polarized tip (e) and a non-spin-polarized tip (f) on the side atom of the Fe~trimer (indicated as small white arrows in (c) and (d)). The tunnel junction setpoint is 54~nS at +15 mV and the external field is 1~T. (g) and (h) $\text{d}I/\text{d}V$ spectra with a spin-polarized tip (g) and a non-spin-polarized tip (h) recorded simultaneously with the $I(V)$ curves shown in (e) and (f). \label{fig1}}
 \end{figure}

The experimental setup is depicted in Fig.~\ref{fig1}a. The few-atom  spin-chain is sandwiched between the tip of a STM and a Cu substrate (Fig.~\ref{fig1}a). The Cu surface was passivated by a monolayer of copper nitride (Cu$_2$N) prior to the experiment and the  spin-chain was assembled by placing three Fe atoms (Fe~trimer) at a spacing of 0.72 nm on Cu binding sites of the Cu$_2$N layer \cite{Loth2010, Loth2012,Yan2015, Yan2015a}. A spin polarization  of the STM tip  of $\eta_t = 0.6$ was achieved by picking up several Fe atoms which remain at the tip apex and applying an external magnetic field of up to 2~T \cite{Bode2003, Wiesendanger2009, SM}. This geometry effectively creates a double-barrier structure into which the Fe~trimer is embedded with one spin-polarized and one non-polarized electrode (Fig.~\ref{fig1}b). At low temperature (0.5 K) this enables long lifetimes of spin excitations in the Fe~trimer, ranging up to microseconds \cite{Yan2015}, such that  tunnel currents as low as picoampere will lead to non-equilibrium transport conditions in this spin system.
\par

With a spin-polarized tip, the $I(V)$ curve recorded above a side atom of the Fe~trimer shows ohmic behavior at small bias and a pronounced non-linearity at  $-$6.7~mV, Fig.~\ref{fig1}e. Between  $-$6~mV and  $-$7.5~mV the current drops with increasing bias amplitude. This clear signature of NDR leads to a pronounced dip in the differential conductance ($\text{d}I/\text{d}V$) spectrum, Fig.~\ref{fig1}g. At positive bias a similar, but weaker, dip is observed at +6.7~mV.
\par

 \begin{figure}
 \includegraphics{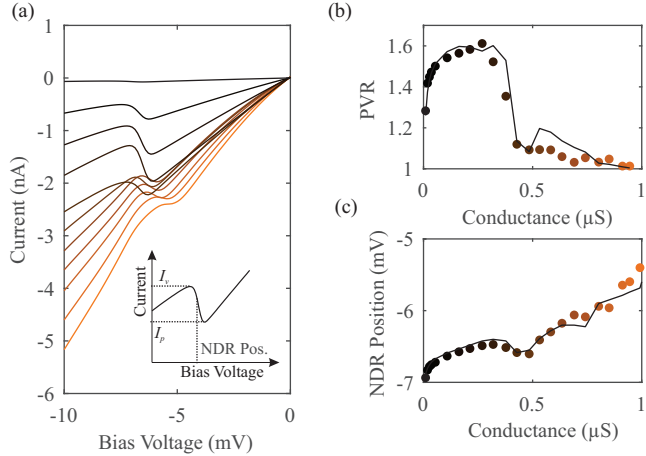}
 \caption{ (a) Conductance-dependent  $I(V)$ curves recorded with a spin-polarized tip on the side atom of the Fe~trimer (tunnel junction conductance changes from 0.01 $\mu$S to 1.06 $\mu$S with about 0.105 $\mu$S per interval and 1~T magnetic field). Inset shows the location of the valley and the peak current ($I_{v}$, $I_{p}$). The position of the NDR is defined as the voltage of highest negative slope of the $I(V)$. (b) Peak-to-valley ratio ($\text{PVR}= I_{p}/I_{v}$) as a function of setpoint conductance (colored points: experimental data; solid line: calculation). (c) Position of the NDR as a function of setpoint conductance (colored points: experimental data; solid line: calculation). \label{fig2}}
 \end{figure}

By contrast, the $I(V)$ curve, recorded with a non-spin-polarized tip is more symmetric on the positive and negative bias sides, and shows no NDR, Fig.~\ref{fig1}f. Instead, it features kinks at +6.8~mV and $-$6.8~mV, which result in steps in $\text{d}I/\text{d}V$ spectrum, Fig.~\ref{fig1}h. The absence of NDR for non-spin-polarized current demonstrates that the NDR is not caused by moving a localized electronic state through the transport channel as found for tunneling through molecules \cite{Repp2005}, non-magnetic clusters or dopants \cite{Rashidi2016}. It must instead be caused by a spin-dependent effect which links to the magnetic states of the Fe~trimer.
\par

The steps in the $\text{d}I/\text{d}V$ spectrum recorded with a non-spin-polarized-tip indicate the onset of significant inelastic tunneling at 6.8~meV that can promote the Fe~trimer into excited spin states \cite{Heinrich2004,Hirjibehedin2006}. This is coincident with the voltage position of the NDR in the spin-polarized $\text{d}I/\text{d}V$ spectrum. To determine whether the observed NDR is indeed caused by dynamic processes \cite{Loth2010,Baumann2015} induced by inelastic electron transport, we performed conductance-dependent measurements with the spin-polarized tip. $I(V)$ curves were recorded with successively increasing setpoint conductance, Fig.~\ref{fig2}a. We characterize the strength of the NDR feature by the peak-to-valley current ratio (PVR) \cite{Grundmann2006}, as shown in Fig.~\ref{fig2}a. Starting at a junction conductance of  0.01~$\mu$S (referenced at +15 mV) we find that the magnitude of the NDR grows as the conductance increases up to a value of 0.27~ $\mu$S, achieving a maximum  value PVR~=~1.61. For even larger conductances, the PVR breaks down at 0.43~$\mu$S and diminishes until the NDR feature vanishes at a conductance of 0.95~$\mu$S, Fig.~\ref{fig2}b. 
\par

These conductance-dependent measurements clearly link the NDR to a dynamic effect induced by the tunneling electrons with energy close to or at the inelastic tunneling threshold. It is therefore likely that it relates to the dynamic interplay between the spin state occupation of the Fe~trimer and the rate at which spin-polarized electrons tunnel through the Fe~trimer. 
\par

To develop a deeper insight, we model the experiment using a Pauli Master equation approach  \cite{Loth2010, Loth2010a, Novaes2010, Ternes2015, Khajetoorians2013, Heinrich2013, Heinrich2015, Delgado2010a, Muenks2016, Wei2013, Xie2013, Zhang2016}, accounting for the average spin state occupation of the Fe trimer and the rate of tunneling electrons between the electrodes (STM tip and Cu substrate) and the Fe~trimer. The time evolution of the spin state population is given as:
\par

\begin{equation}
\frac{\text{d}n_{i}(t)}{\text{d}t}=\sum_{j} [{ r_{ij} n_{j}(t) -   r_{ji} n_{i}(t) }]
\label{eq1}
\end{equation}

where $n_i (t)$ is the average occupation of each spin state and $r_{ij}$ is the transition rate from state $j$ to $i$. The transition rate $r_{ij} =  r_{ij}^{s \leftarrow 1 \leftarrow t}+ r_{ij}^{t \leftarrow 1 \leftarrow  s}+  r_{ij}^{t \leftarrow 1 \leftarrow  t} + \sum_{a=1}^3 r_{ij}^{s \leftarrow a \leftarrow  s}  $ is  the sum of all possible transition rates causing the transition from $j$ to $i$. The possible rates are the tunneling events through the side atom $"1"$ from STM tip ($t$) to the Cu substrate ($s$) (${s \leftarrow 1 \leftarrow  t}$) and vice versa (${t \leftarrow 1 \leftarrow  s}$), as well as backscattering into the tip  (${t \leftarrow 1 \leftarrow  t}$), and from substrate back into the substrate (${s \leftarrow a  \leftarrow  s}$) through each of the three Fe atoms  ($a=1,2,3$) of the Fe trimer. 
\par

The model links the transition rates of tunneling electrons between STM tip and substrate directly to the rate of transitions between spin states of the Fe~trimer. This link originates from the nature of the electron-spin scattering that is well-described as Kondo-type electron spin scattering \cite{Appelbaum1967,Delgado2010,Ternes2015}. A scattering Hamiltonian of the form $ \hat{H}_s = \hat{\vec{S}} \hat{\vec{\sigma}} + u \hat{I}$ is used, where $\hat{\vec{S}} $ is the vector spin operator of the magnetic atom with which the electron interacts, $\hat{\vec{\sigma}}$ is the vector spin operator of the tunneling electron and $\hat{I}$ is the identity operator accounting for spin-independent scattering with strength $u$. This scattering Hamiltonian was previously found to quantitatively account for elastic and inelastic electron tunneling at magnetic atoms and molecules \cite{Kim2004,Fernandez-Rossier2009,Lorente2009, Loth2010, Ternes2017, Spinelli2014,Bryant2013,Kahle2012, Hurley2011,Gauyacq2013}. To first order, the transiton rates are given by the product of this operator’s transition matrix elements between the spin states of Fe~trimer $|i\rangle$ and $|j\rangle$ \cite{SM}.
\par

\begin{figure}
 \includegraphics{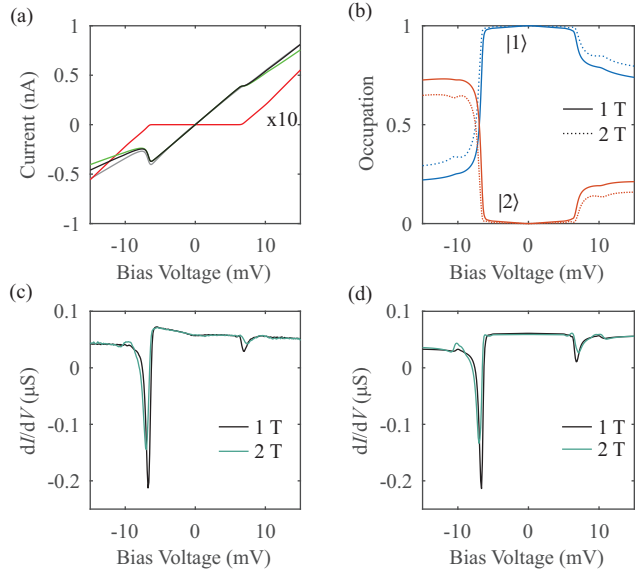}
 \caption{ (a) Calculated (black) and measured (grey) spin polarized $I(V)$ curves on the side atom (setpoint of 0.054~$\mu$S at 1~T). The elastic (green) and the inelastic (red, 10x magnified) current contributions are shown separately. (b) Calculated voltage-dependent occupation, $n_i(V)$, of the two low-energy states ($|1\rangle$ and $|2\rangle$) for 1~T (solid) and 2~T (dashed) magnetic field. (c) The $\text{d}I/\text{d}V$ spectra measured at 1 T (black) and 2~ T (green) with setpoint of 0.054 $\mu$S. These two spectra are measured with the same spin-polarized tip on the same side atom of Fe trimer. (d) The correspondingly calculated $\text{d}I/\text{d}V$ spectra at 1~T (black) and 2~T (green) with setpoint of 0.054~$\mu$S. \label{fig3}}
 \end{figure}

Then, the $I(V)$ curve can be expressed as:

\begin{equation}
I(V)= e \sum_{j,i} [ r_{ij}^{s \leftarrow 1 \leftarrow t}(V) - r_{ij}^{t \leftarrow  1 \leftarrow s}(V)]   \tilde{n}_j (V) \label{eq2}
\end{equation}
where  $\tilde{n}_j(V)$ are the steady state solutions of equation (\ref{eq1}). Taking the numerical derivative of the $I(V)$ curves yields the $\text{d}I/\text{d}V$ spectra which are fit to the experimental data using least squares optimization \cite{SM}. Figures~\ref{fig3}a,c,d show that calculated $I(V)$ curves and $\text{d}I/\text{d}V$ spectra reproduce the experimental data quantitatively. The calculated tunneling current is the sum of inelastic transitions, $i\neq j $, and elastic transitions, $i=j$. We find that the inelastic tunnel current has a fast onset at bias $|V|~>~6.4$~mV and rises monotonically with increasing bias magnitude, Fig.~\ref{fig3}a. The NDR stems from the elastic current that drops sharply with the onset of inelastic tunneling, Fig.~\ref{fig3}a. 
\par

This behavior can be understood by the dynamic reconfiguration of the Fe trimer's spin state occupation when tunnel current passes through it. The Fe~trimer features two low-energy spin states. They are the Néel-like antiferromagnetic configurations of the Fe atom magnetic moments and denoted as $|1\rangle$ and $|2\rangle$ in the following. Due to the odd number of magnetic atoms these states split in magnetic field by 0.46 meV/T \cite{Yan2015}.
\par

At our measurement conditions (0.5 K temperature,  1~T or 2~T magnetic field), thermal occupation of state $|2\rangle$ is negligible. As the sample bias decreases from zero, the occupation of state $|2\rangle$ increases only slowly since the probability for direct transition between the two states is small. All electric current is carried by elastic spin-dependent tunneling.
When the bias drops below $-$6.8~mV, state $|2\rangle$ becomes populated rapidly and dynamically locked in this state, Fig.~\ref{fig3}b. State $|2\rangle$  has a markedly smaller transition matrix element for elastic tunneling than state $|1\rangle$ because of the spin-polarized STM tip. Hence the rapid occupation change decreases the conductance quickly and the net current through the Fe~trimer drops. The effective magnetoresistance between spin-polarized STM tip and Fe trimer is therefore bias-dependent and causes the NDR.
\par

Our calculations show that the occupation change at $-$6.8~mV is driven by an efficient two-step transition through intermediate states where the spin momentum of the Fe trimer is decreased by $1\hbar$. The intermediate states, located 6.8~meV above the ground state, are energetically unfavorable and they do not become significantly populated because of magnetic anisotropy and inter-atomic exchange energy of the Fe trimer \cite{SM}.
\par

The presence of this process is corroborated by the asymmetry of the $I(V)$ curves with respect to bias reversal and their variation with magnetic field. 
Due to spin-momentum conservation tunneling electrons exciting the two-step transition must change their spin by $+1\hbar$, i.e., from $\langle \sigma_z \rangle=-1/2 $ to $\langle \sigma_z \rangle=+1/2 $. With negative bias, electrons tunnel into $\sigma_z =+1/2$ states of the tip  whereas they tunnel  out of $ \sigma_z =-1/2$ states for positive bias since the tip is strongly spin polarized. Hence, inelastic transitions are more frequent at negative bias and generate a more pronounced dip in $\text{d}I/\text{d}V$. In addition, the transition energy between the intermediate states and states $|1\rangle$ and $|2\rangle$ depends on the magnetic field because of the difference in their net magnetic moment. Upon increasing the magnetic field from 1~T to 2~T the NDR feature shifts to higher energy, consistent with the difference in Zeeman energy between the states (Figs.~\ref{fig3}c,d). Surprisingly the PVR, i.e., the NDR amplitude, also reduces. This is caused by an increase of the transition rate from state $|2\rangle$ into the intermediate  states which reduces the occupation of $|2\rangle$ compared to 1~T, Fig.~\ref{fig3}b.
\par

This sensitivity of the NDR to magnetic fields points to the importance of considering the coupling between the spin-polarized tip and the Fe trimer in more detail. For tunnel current to flow, wave functions of the Fe trimer and the tip must overlap. Since the tip apex is spin-polarized and magnetic, the same overlap may induce magnetic exchange interaction \cite{Ferriani2010}.
Indeed, the reduction of the PVR for junction conductances in excess of 0.27~$\mu$S (Fig.~\ref{fig2}b) cannot be explained by the increasing tunneling rates alone. Antiferromagnetic interaction between spin-polarized STM tip and Fe~trimer must be included in order achieve quantitative fits to the $I(V)$ curves \cite{SM}.
\par

\begin{figure}
 \includegraphics{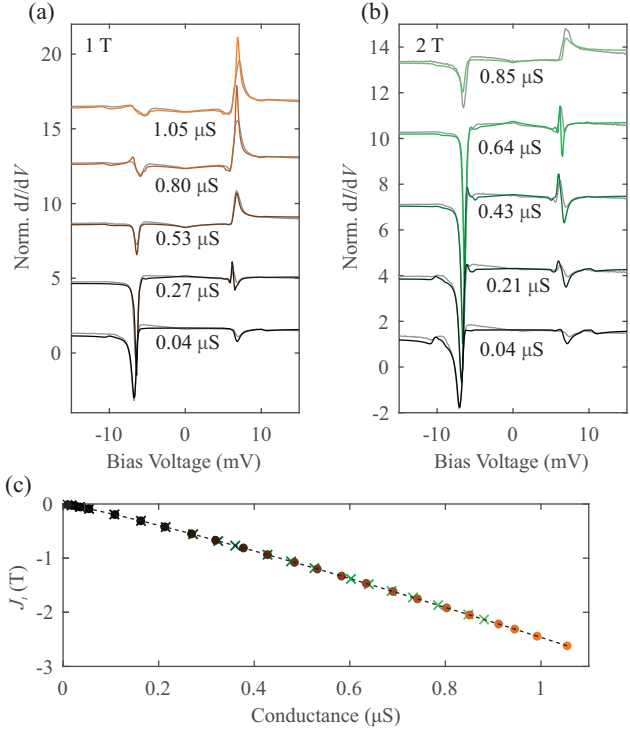}
 \caption{(a) and (b) The  measured conductance-dependent  $\text{d}I/\text{d}V$ curves (grey lines) with 1~T and 2~T magnetic field applied, respectively. The calculated $I(V)$ curves are shown as colored lines. For visibility the curves are normalized the setpoint conductance and are offset. (b) The extracted strength of the tip interaction as function of the conductance setpoint by fitting the measured conductance-dependent $\text{d}I/\text{d}V$ spectra. Colored markers indicate the junction conducances used for the fit (dots: 1~T, crosses: 2~T). \label{fig4}}
 \end{figure}

The interaction reduces the energy separation between states $|1\rangle$ and $|2\rangle$ and increases the transition probability. For an applied magentic field of 1~T and junction conductances larger than 0.27~$\mu$S this leads to significant population of state $|2\rangle$ even at $|V| < 6.8$~mV (0.5 K). Therefore, upon reaching the excitation threshold, the change in the state population is less pronounced compared to smaller junction conductance and the NDR is reduced (reduced PVR in Fig.~\ref{fig2}b).
In addition, the energy separation to the intermediate spin state of the two-step excitation reduces which is observed as a position shift of the NDR to lower bias, Fig.~\ref{fig2}c. The center of the NDR shifts from $-$6.8 mV for junction conductance of 0.01~$\mu$S to $-$5.6~mV at conductance of 0.95~$\mu$S.
Notably, at junction conductance of 0.43~$\mu$S states $|1\rangle$ and $|2\rangle$ reverse order because the coupling with the spin-polarized tip becomes stronger than the external magnetic field and all dynamic processes reverse: the magnetoresistance for elastic tunneling is now large at small bias and does not increase further when the bias magnitude exceeds 5.4~mV.  Consequently, the NDR at negative bias disappears and a pronounced conductance peak appears at positive bias, Fig.~\ref{fig4}a. Upon increasing the magnetic field to 2~T, the conductance dependent $\text{d}I/\text{d}V$ spectra change slightly (Fig.~\ref{fig4}a and Fig.~\ref{fig4}b). For 2~T magnetic field, a stronger coupling strength with the spin-polarized tip is needed to reverse the dynamic processes. For that reason the conductance dip on the negative bias side disappears  at larger junction conductance in comparison to  1~T external magnetic field (Fig.~\ref{fig4}a and Fig.~\ref{fig4}b). 

\par

Since tunnel current and exchange interaction typically decay exponentially with tunneling gap size \cite{Yan2015, Schmidt2011} we consider a conductance-dependent Ising interaction of the form $J_t(\sigma)\sim \sigma^\kappa$, where $\kappa$ is the ratio of decay constants for tunneling and magnetic interaction. By fitting the conductance dependent $\text{d}I/\text{d}V$ spectra with 1~T and 2~T magnetic field simultaneously, we find $\kappa=(1.1 \pm 0.1)$ reproduces both measurements with high accuracy, Fig.~\ref{fig4}c.
Hence, magnetic interaction and tunnel current are carried by wave functions with similar decay into vacuum emphasizing that both effects are of equal importance for the spin-based NDR.
\par
\par

In conclusion, we report spin-based NDR from a antiferromagnetically coupled few-atom  spin-chain in a spin-polarized tunneling junction. The NDR occurs at low bias magnitude of 7~mV and is caused by a two-step excitation that drives fast transitions between the two Neel-like spin states of the  spin-chain and dynamically locks it in a configuration with large magnetoresistance. The NDR varies non-monotonically with the junction conductance between spin-polarized tip and spin-chain. We attribute this behavior to a competition between dynamic locking by inelastic electron tunneling that produces NDR and static magnetic exchange interaction that prevents it. Our measurements suggest that both effects are mediated by similar wave functions and likely linked inseparably. 
This spin-based NDR will be observable for other magnetic nano-objects with discrete spin states, such as single molecule magnets \cite{Elste2006} and nano-ferromagnets \cite{Spinelli2014}. As it occurs at very low voltages, on the order of the magnetic anisotropy and at low currents, it may prove useful for applications in low-current spintronic devices. Due to the sensitivity of this NDR to magnetic interaction the effect has potential as the basis for an atomic-scale magnetic sensor.

\begin{acknowledgments}
The authors thank E. Weckert and H. Dosch (Deutsches Elektronen-Synchrotron, Germany) for providing lab space. This project has received funding from the European Research Council (ERC) under the European Union's Horizon 2020 research and innovation programme (ERC-2014-StG-633818-dasQ). SRP acknowledges a fellowship from the German Academic Scholarship Foundation and JAJB postdoctoral fellowships from the Alexander von Humboldt foundation and the Natural Sciences and Engineering Research Council of Canada.
\end{acknowledgments}

\bibliography{NDR}

\end{document}